\documentclass[nofootinbib,aps,twocolumn,letterpaper,preprintnumbers,superscriptaddress]{revtex4-2} 
\usepackage{amsmath}
\usepackage{eurosym}
\usepackage{amssymb}
\usepackage{graphicx,epsfig}

\usepackage{color}
\usepackage{braket}
\usepackage{booktabs}
\usepackage{dcolumn}
\usepackage{multirow}
\usepackage{amsmath}
\usepackage{amssymb}
\usepackage{subfigure}
\usepackage{amsfonts}
\usepackage[left=1.5cm,right=1.5cm,top=1.5cm,bottom=1.5cm]{geometry}
\usepackage[usenames,dvipsnames,svgnames]{xcolor} %% colored text
\usepackage{hyperref}
\definecolor{darkgreen}{rgb}{0.1,0.6,0.1}
\definecolor{darkblue}{rgb}{0,0,0.3}
\definecolor{darkred}{rgb}{0.7,0,0}

\definecolor{light gray}{RGB}{220,220,220}
\definecolor{dark purple}{RGB}{108,0,217}
\definecolor{pink}{RGB}{190,20,100}
\definecolor{orang}{RGB}{193,63,0}
\definecolor{green}{RGB}{11,98,17}
\definecolor{darkpink}{RGB}{153,0,76}
\definecolor{bluegreen}{RGB}{0,102,102}
\definecolor{greenlagan}{RGB}{0,102,0}
\definecolor{redgreen}{RGB}{102,102,0}
\definecolor{Redgreen}{RGB}{153,76,0}
\definecolor{vividviolet}{rgb}{0.62, 0.0, 1.0}
\definecolor{amaranth}{rgb}{0.9, 0.17, 0.31}
\definecolor{palatinateblue}{rgb}{0.15, 0.23, 0.89}
\definecolor{brightpink}{rgb}{1.0, 0.0, 0.5}
\definecolor{cornflowerblue}{rgb}{0.39, 0.58, 0.93}
\definecolor{deepcarminepink}{rgb}{0.94, 0.19, 0.22}
\definecolor{radicalred}{rgb}{1.0, 0.21, 0.37}

\definecolor{beamer@PRD}{RGB}{46,48,146}
\hypersetup{
breaklinks=true,
pdfstartview={FitH}, % fits the width of the page to the window
colorlinks=true, % false: boxed links; true: colored links
linkcolor=vividviolet, % color of internal links
citecolor=blue, % color of links to bibliography
filecolor=magenta, % color of file links
urlcolor=brightpink, % color of external links
anchorcolor=cyan, % Color for anchor text
linktocpage=true
}

{\vskip .1cm}
\date{\today}
\newcommand\be{\begin{equation}}
\newcommand\ee{\end{equation}}
\newcommand\bea{\begin{eqnarray}}
\newcommand\eea{\end{eqnarray}}
\newcommand\bseq{\begin{subequations}} %solo con amsmath
\newcommand\eseq{\end{subequations}}
\newcommand\bcas{\begin{cases}}
\newcommand\ecas{\end{cases}}

\begin{document}
\title{Growth of perturbations in higher dimensional Gauss-Bonnet FRW cosmology}

\author{\bf Ahmad Sheykhi}
\email{asheykhi@shirazu.ac.ir}
\affiliation{Department of Physics, College of Sciences, Shiraz University, Shiraz 71454, Iran}
\affiliation{Biruni Observatory, College of Sciences, Shiraz University, Shiraz 71454, Iran}

\author{\bf Bita Farsi}
\email{Bita.Farsi@shirazu.ac.ir}
\affiliation{Department of
Physics, College of Sciences, Shiraz University, Shiraz 71454,
Iran}

%%%%%%%%%%%%%%%%%%%%%%%%%%%%%%%

\begin{abstract}
We explore the influences of the higher-order Gauss-Bonnet (GB)
correction terms on the growth of perturbations at the early stage
of a $(n+1)$-dimensional Friedmann-Robertson-Walker (FRW) universe. 
Considering a cosmological constant in the FRW background, we study the
linear perturbations by adopting the spherically symmetric
collapse (SC) formalism. In light of the modifications that appear in the field equations, we disclose the role of the GB coupling constant $\alpha$, as well as the extra dimensions $n>3$ on the growth of perturbations. 
It, in essence, is done by defining a dimensionless parameter $\tilde{\beta}=(n-2)(n-3) \alpha H_0^2$ 
in which $H_0$ is the Hubble constant. We find that the matter density contrast starts
growing at the early stages of the universe and, as the universe
expands, it grows faster compared to the standard cosmology.
Besides, in the framework of GB gravity, the growth of matter perturbations in higher dimensions is faster than its standard counterpart $(n=3)$.
Further, in the presence of $\alpha$, the growth of perturbations increases as it increases. This is an expected result, since
the higher order GB correction terms increase the strength of the
gravity and thus support the growth of perturbations. For the existing cosmological model, we also investigate the behavior of quantities such as density abundance, deceleration, and the jerk parameter. Finally, we study the imprint of  the GB parameter and the higher dimensions in the evolution of the mass function of the dark matter halos.

\end{abstract}
\maketitle

 %%%%%%%%%%%%%%%%%%%%%%%%%%%%%%%%%%%%%%%%%%%%%%%%%%%%%%%%%%%%%%%%%%%%%%%%%%%%%%%%%%%
\section{Introduction \label{Intro}}
It is a general belief that general relativity (GR) is the most
prosperous gravity theory for describing the physical and
cosmic phenomena over a wide range of energies from large
scales to small scales \cite{Will}. However, it is not a perfect
theory based on observational limitations and theoretical
considerations. Specifically, two dark clouds of modern physics,
dark matter (DM) \cite{Young} and dark energy (DE) \cite{Weinberg,
Riess, Perlmutter} can not be well explained in the framework of
GR plus $\Lambda$-cold DM ($\Lambda$CDM). This implies that the
underlying gravity theory governing the gravitational dynamics of
the universe may not be GR and could be an alternative
gravitational scenario, which can help understand the dark sector
better at least. This motivates physicists to pay attention to
modified theories of gravity. Although the motivations behind these alternatives to GR may come from mathematics, philosophy, and observation, almost all lead to generalizations of Einstein's theory. Among all these possibilities Einstein-Gauss-Bonnet (EGB) gravity
which initially proposed by Lanczos \cite{Lanczos, Lanczos2},
and subsequently generalized by Lovelock \cite{Lovelock, Lovelock
2} has a special place. Lovelock's theorem, in essence, lets us find all possible field equations that have consistent conservation and symmetry properties, just like Einstein’s field equations. Features that make EGB a plausible candidate for generalizing GR include the ghost-free gravitation propagator \cite{Zumino00},
natural generalization with Einstein, and cosmological terms \cite {Zwiebach}. The presence of the GB term leads to second-order field equations, which are free of  Ostrogradski instabilities since it is a unique term that is quadratic in the curvature.
It is intriguing to note that EGB theory is backed by a fundamental theory, namely string theory, meaning that at the classical limit, Einstein's equations are subject to leading-order corrections, which typically arise from higher-order curvature terms in the action.  
Based on Lovelock's theorem, the GB term should appear beyond four dimensions since Einstein's equations in four dimensions are the most general set of field equations that satisfy all conditions proposed in Lovelock's theorem. It means that to find the contribution of the GB term in cosmology, one must explore it in higher-dimensional cosmology \footnote{It is important to note one comment here. Recently, Glavan and Lin, in an attempt to directly introduce the GB term in four-dimensional gravity, have proposed a covariant-modified gravity well-known as 4DEGB, which bypasses the requirements of Lovelock's theorem \cite{Glavan}. Although our focus throughout this manuscript is on the role of extra dimensions GB, it can be helpful to mention a few of the research done on some aspects of 4DEGB theory that in recent years have been investigated in the framework of cosmology \cite{Wu,Aoki,Sadjadi,Narain,Feng,Miguel,Aoki2,Wang, Shahidi} and 		the black hole (compact objects) physics \cite{Kumar,Zhang,Aragon,Yang,Lin,Wei,Konoplya,Konoplya2,Yang2,Heydari-Fard,Hosseini Mansoori,Wei2}. In this regard, a review paper has also been written that provides a comprehensive discussion of 4DEGB \cite{Fernandes:2022zrq}.} \cite{Lorenz, Andrew, Atmjeet}. 

Could there be additional spatial dimensions beyond the three that we know? Theoretically, there is no concrete a priori reason that forbids us to pursue this fundamental question (for overviews of diverse theories with extra dimensions and their physical consequences see e.g.,\cite{Overduin,Rubakov,Brax}). One of the well-known reasons for this is that phenomena that need very different justifications in three-dimensional space are expected to be nothing more than manifestations of simpler theories in higher-dimensional manifolds \cite{Overduin}. But the main issue is how to adapt this idea to the real universe i.e., the three-dimensionality of space. 
Both our daily experience and the experiments of particle and space physics clearly show that we live in a four-dimensional universe-three for space and one for time. Besides, in light of the joint detection of gravitational waves and electromagnetic signals \cite{Visinelli:2017bny} and recent Event Horizon Telescope data \cite{Vagnozzi:2019apd}, the possibility of the extra dimension(s) in the real universe can not be ruled out.
Therefore, to justify the extra dimensions and search for them, we have to explain where are these extra dimensions. The commonly answer is that the extra dimensions are compactified one a very small scale. 
But this answer, is not the end of the story and gives rise to another question: how come they became compact? The answer to this question is not straightforward. A solution known as spontaneous compactification is the first attempt to handle this fundamental issue \cite{Hoissen,Deruelle}. In Refs. \cite{Hoissen2, Pavluchenko}, one can find similar solutions with approaches closer to cosmology.  In this direction, a more natural way to achieve compactified extra dimensions is dynamical cementification (see Refs. \cite{Mena, Elizalde0, Maeda, Maeda2} involving different approaches and setups).

It is well-known for a long time that extra-dimensional theories can, in the appropriate limit, behave like a standard four-dimensional spacetime with additional field content derived from the imprint of the extra dimensions. Extra dimensions have particular applications and consequences in the context of cosmology, including questions on the nature of DM and DE. Theories of matter existing in hidden extra dimensions, or the extra-dimensional effects on the dynamics of the observed four-dimensional universe could potentially shed light on these issues. Similarly, early-universe phenomena such as inflation could be driven by extra-dimensional effects, or at least are allowed to occur in the presence of extra-dimensions
\cite{Kanti,Wongjun,Nojiri,Amendola,Amendola2,Koivisto,Neupane,Leith,
Odintsov0}. Considering the extra-dimensional GB in cosmology can be a source for the generation of primordial cosmic magnetic fields \cite{Atmjeet:2013yta}. Recently the GB cosmology with extra spacetime
dimensions has been investigated \cite{Chirkov0}. Some aspects of
the dynamical compactification scenario, where stabilization of
extra dimensions occurs due to presence the GB term and non-zero
spatial curvature, have been explored in \cite{Chirkov2}.

For all mentioned above, it becomes obvious that investigation the
structure formation at the early stages of the universe in the
context of GB cosmology with extra dimensions is well motivated.
In this work we disclose the effects of the GB correction terms on
the gravity side, as well as the extra dimensions on the growth of
perturbations at the early stages of the universe. Since the GB
correction terms contribute to the dynamical field equations in
more than four spacetime dimensions, thus, in order to explore the
role of the GB correction terms on the growth of perturbation, we
need to consider higher spacetime dimensions. We shall use the SC
formalism \cite{Abramo} to examine the growth of perturbations and
structure formation. In this approach one considers a uniform and
spherical symmetric perturbation in an expanding background and
describes the growth of perturbations in a spherical region using
the same Friedmann equations for the underlying theory of gravity
\cite{Planelles,Ziaie,Ziaie2,Farsi,Farsi2}.

The outline of this paper is as follows. In Sec. II, we provide a
review on GB gravity and derive the corresponding Friedmann
equations in the context of (n+1)-dimensional GB cosmology. In
Sec. III, using the spherically collapse approach, we explore the
growth of matter perturbation in the background of the GB
cosmology in higher dimensions. In section IV, we investigate the
effects of the GB parameter and extra dimensions on the mass
function of the dark matter halos. We finish with conclusions and
discussions in the last section.
%%%%%%%%%%%%%%%%%%%%%%%%%%%%%%%%%%%%%%%%%%%%%%%%%%%%%%%%%%%%%%%%%%%%%%%%%
 \section{Modified Friedmann Equations higher-dimensional GB Cosmology\label{FGC}}
The action of the GB gravity in $(n+1)$-dimensional spacetime, and
in the presence of cosmological constant $\Lambda$, can be written
as \cite{CaiGB}
\begin{eqnarray}\label{action}
S_{EGB}=\frac{1}{2 \kappa_{n+1}^{2}}\int{d^{n+1}x
\sqrt{-g}\left(R-2\Lambda+\alpha \mathcal{L}_{GB}\right)}+S_{m},
\end{eqnarray}
where $\alpha$ is called the GB coupling constant which has
the dimension $[\alpha]=[length]^{2}$, $\mathcal{L}_{GB}=R^{2}-4R_{\mu
\nu}R^{\mu \nu}+R_{\mu \nu \gamma \delta} R^{\mu \nu \gamma
\delta}$ is the GB Lagrangian, and $S_{m}$ denotes the action of
matter. In light of the string theory evaluations, it is well known that the GB coupling constant should be positive, i.e., $\alpha>0$ (see \cite{Boulware:1985wk,Paul:1990jg}).
The field equations can be derived by varying the above
action with respect to the metric. One finds 
\begin{eqnarray}
&&\kappa_{n+1}^{2} T_{\mu \nu} = R_{\mu \nu}-\dfrac{1}{2} g_{\mu \nu}R+\Lambda g_{\mu \nu}
-\alpha \Big{\{} \dfrac{1}{2}g_{\mu \nu}\mathcal{L}_{GB}\nonumber\\
&&-2RR_{\mu \nu}+4R_{\mu \gamma}R^{\gamma}_{\   \nu} +4R_{\gamma
\delta}R^{\gamma \  \delta}_{\ \mu \  \nu}-2R_{\mu \gamma \delta
\lambda} R_{\nu}^{\     \gamma \delta \lambda}\Big{\}}.\nonumber\\
\label{field}
\end{eqnarray}
In what follows we work in the units where
$\hslash=c=\kappa_{n+1}=1$. The line elements of a spatially flat $(n+1)$-dimensional Friedmann-Robertson-Walker (FRW) metric for the homogeneous and isotropic universe is as follows \footnote{For the sake of simplicity, we employ a higher dimensional FRW metric with the assumption that the extra spatial dimensions are also homogeneous and isotropic. In this direction, it can be helpful to remember that exists another cosmological setting based on Kaluza-Klein theory which considers a spacetime with manifold $M^4\times T^{d=n-3}$. The line element of this manifold, is given by
\begin{eqnarray}
ds^2=-dt^2+a^2(t)\left(\frac{dr^2}{1-Kr^2}+r^2d \Omega^2\right)+
b(t)^2\gamma_{ab}dy^ady^b,~~~K=0,\pm1 \nonumber
\end{eqnarray}
In this ansatz metric, there is a generalized term in the form $b(t)^2\gamma_{ab}dy^ady^b$, in addition to the spatial manifold labeled by the coordinates $(r,\Omega)$. It  includes the extra spatial coordinates $y^{a}$, the compact manifold described by the metric $\gamma_{ab}$, and the scale factor of extra dimensions $b(t)$. Overall, the underlying spacetime is no longer isotropic since $a(t)\neq b(t)$. The underlying compact manifold is assumed to be maximally symmetric whose Riemann tensor for the metric $\gamma_{ab}$ is read as $R_{abcd}=k(\gamma_{ac} \gamma_{bd}-\gamma_{ad}\gamma_{bc})$. Despite the absence of observational support for the flatness spatial curvature of the extra dimensions, interestingly, in Ref. \cite{Mohammedi:2002tv} (see also \cite{Andrew}), considering the contribution of the compact manifold corresponding to the extra dimensions in the metric, it is shown that it should be flat (i.e., $k=0$) to avoid some unphysical properties of energy density and perfect fluid pressure. It means that, in case of extending the assumption of flatness spatial curvature to higher dimensions, no conflict will arise.}
\begin{equation}\label{metric}
 ds^2=-dt^2+a^2(t)\left(dr^2+r^2d \Omega^2_{n-1}\right),
\end{equation} where $d \Omega^2_{n-1}=d\theta_1^2+ Sin^2\theta_1d\theta_2^2...+ Sin^2\theta_{n-2}d\theta_{n-1}^2$, represents the line element of a $(n-1)$-dimensional sphere.
Substituting metric (\ref{metric}) in the gravitational field
equations (\ref{field}), and assuming the matter content of the
universe is in the form of perfect fluid, one get the
corresponding Friedmann equations as \cite{Cai}
\begin{eqnarray}\label{Fried1}
&& H^{2}+\tilde{\alpha}H^{4}=\dfrac{2}{n(n-1)}\left(
\rho_{m}+\Lambda \right),\\
&&\left(1+2 \tilde{\alpha}H^{2}\right) \dot{H}
=-\dfrac{1}{n-1}\rho_{m}, \label{Fried2}
\end{eqnarray}
where $\tilde{\alpha}=(n-2)(n-3) \alpha$, $H\equiv \dot{a}/a$ is
the Hubble parameter, $\rho_{m}$ is the energy density of both
baryonic and DM. Notice that in the limiting case where $n=3,
\tilde{\alpha}=0 $, Eqs. (\ref{Fried1}) and (\ref{Fried2}) reduce to the
Friedmann equations in standard cosmology, meaning that in four dimensions the GB coupling constant does not contribute to the dynamic equations. Therefore in four dimensions, $(n=3)$, we should replace $\alpha\rightarrow \alpha / (n-3)$. Of course, in case of coupling to the scalar field is not the case and leaves some non-trivial effects in four-dimension \cite{Amendola,Amendola2}. Concerning the higher dimensions, the cases $n=4$, and $n=5$, are more popular than others, especially  $n=4$, which results in certain mathematical simplifications (e.g., see \cite{Hansraj:2020hwf,Bogadi:2023moe}).
Moreover, the
continuity equation in $(n+1)$-dimensions can be written as
\begin{equation}
\dot{\rho}_{m}+nH(\rho_m+p_m)=0. \label{continuty}
\end{equation}
The energy density of the pressureless matter ($p_m=0$) can be
obtained as $\rho_{m}=\rho_{m,0}a^{-n}$. Therefore, with the
following dimensionless parameters:
%\begin{eqnarray}
%\beta\equiv\alpha H_{0}^{2}, \label{beta}
%\end{eqnarray}
\begin{eqnarray}
\tilde{\beta}\equiv\tilde{\alpha}H_{0}^{2}=(n-2)(n-3)\alpha H_{0}^{2},
\label{betatilde}
\end{eqnarray}
\begin{eqnarray}\label{OMmatter}
&& \Omega_{m}=\frac{2\rho_{m}}{n(n-1)H^{2}},\\ &&
\Omega_{\Lambda}=\frac{2\Lambda}{n(n-1)H^{2}}.\label{OmLambda}
\end{eqnarray}
One can rewrite Eq. (\ref{Fried1}) as
\begin{equation}
E^{2}(z)+ \tilde{\beta} E^{4} (z)= \Omega_{m,0}(1+z)^{n}+\Omega_{\Lambda,0},
\label{FriedEZ}
\end{equation}
where $E(z)=H(z)/H_0$. This equation governs the evolution of the
homogeneous universe in the context of an $(n+1)$-dimensional GB
gravity. Moreover, at the present-time($z=0$), Eq.
(\ref{FriedEZ}), reduces to
\begin{equation}
\Omega_{m,0}+\Omega_{\Lambda,0}=1+\tilde{\beta} .
\label{Omega0}
\end{equation}
Notice that when $\tilde{\beta} \rightarrow 0$, the standard
equation is recovered. By extending the 
flatness assumption to higher dimensions (see Footnote 2), thereby, the relation  
$\Omega_{m}+\Omega_{\Lambda}\simeq1$ holds in all dimensions. 
 This implies that the value of the dimensionless coupling parameter, $\tilde{\beta}$, should be very small. The stringent upper bounds obtained in light of cosmological settings for the dimensionless parameter $\alpha H_{0}^{2}$ also confirm that it is merely a perturbative constant \cite{Amendola,Amendola2}. We do so throughout our analysis.

Solving Eq. (\ref{FriedEZ}) with respect to $E(z)$ at a given
redshift $z$, and considering the branch where we have a real
value of $E(z)$, yields
\begin{equation}
E^{2}(z)=\dfrac{H^{2}(z)}{H_{0}^{2}}=\dfrac{1}{2\tilde{\beta}}\left[ \sqrt{X(z)}-1\right],
\label{Ez}
\end{equation}
where
\begin{equation}
X(z)\equiv1+4 \tilde{\beta} \left[ \Omega_{m,0} (1+z)^n +
\Omega_{\Lambda,0} \right]. \label{Xz}
\end{equation}
In general for any $z$ we have:\\
\begin{equation}
\Omega_{m}(z)+\Omega_{\Lambda}(z)=1+\tilde{\beta}E^{2}(z).
\label{Omegat}
\end{equation}
The Hubble expansion rate can be obtained via Eqs.
(\ref{OMmatter}), (\ref{OmLambda}) ,(\ref{Ez}), (\ref{Xz}). We
find
\begin{eqnarray}  \label{Hz}
&&H^{2}(z)= \dfrac{H_{0}^{2}}{2\tilde{\beta}}\left[
\sqrt{X(z)}-1\right]\nonumber\\ &&=
\dfrac{2}{n(n-1)}(\rho_{m}+\Lambda)\left(1-\dfrac{2\tilde{\alpha}}{n(n-1)}(\rho_{m}+\Lambda)
\right)
\end{eqnarray}
where we have expanded $X(z)$ and only kept the linear term of
$\tilde{\beta}$, since $\tilde{\beta}\ll 1$ is very small.

\begin{figure}[t]
\epsfxsize=7.3cm \centerline{\epsffile{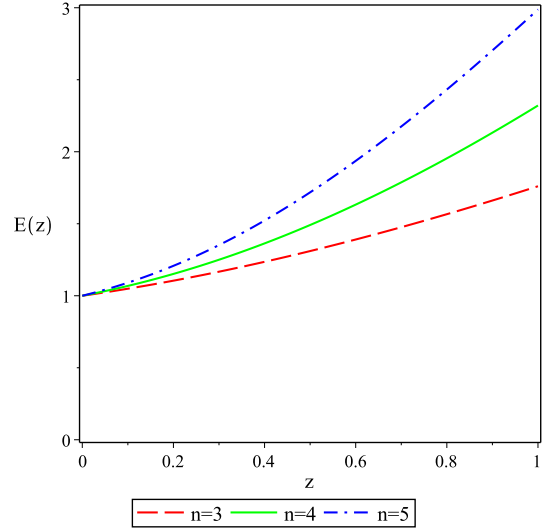}} \caption{The
behavior of the normalized Hubble rate $E(z)$ for different values
of $n$ in GB cosmology, where we have taken $\tilde{\beta}=10^{-16}$.} \label{Fig1}
\end{figure}

\begin{figure}[t]
\epsfxsize=7.3cm \centerline{\epsffile{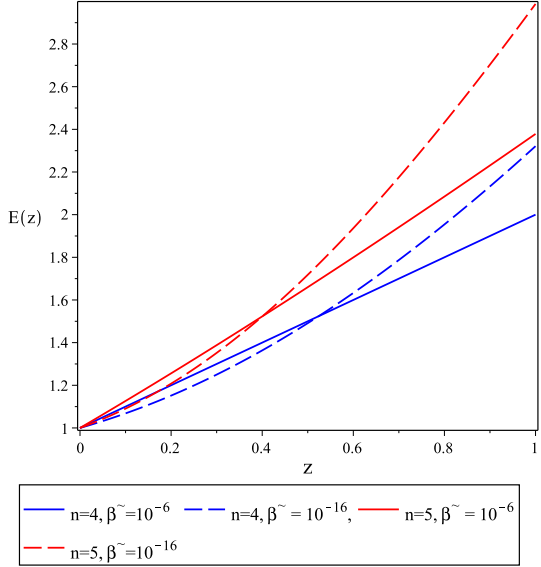}} \caption{The
behavior of the normalized Hubble rate $E(z)$ for different values
of $(n,\tilde{\beta})$.}
\label{Fig2}
\end{figure}

\begin{figure}[t]
\epsfxsize=7.5cm \centerline{\epsffile{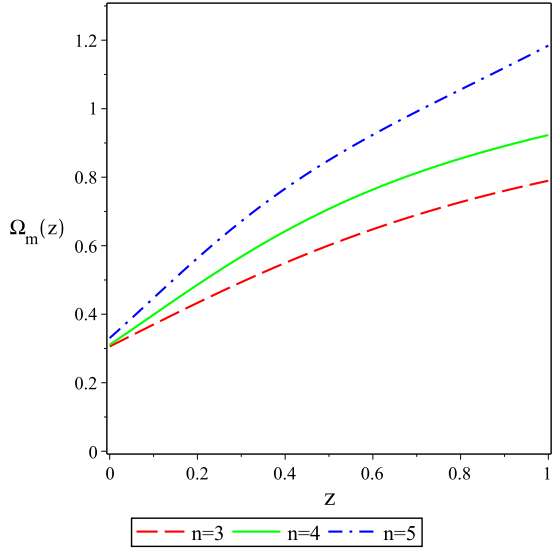}} \caption{The evolution of the matter density abundance as a
function of redshift z for different values of $n$, where we have taken $\tilde{\beta}=10^{-16}$.} \label{Fig3}
\end{figure}

The evolution of the normalized Hubble parameter versus $z$ for
different values of $n$ is plotted in Fig. \ref{Fig1}. As we can see,
in GB cosmology the Hubble parameter with higher dimensions are larger than
lower dimensions model, implying that in lower dimensions model, our Universe
expands slower.
Also we have plotted this normalized Hubble parameter for
different values of $\tilde{\beta}$ and $n$ in Fig. \ref{Fig2}. As we
can see, in GB cosmology the Hubble parameter at high redshifts
decreases with increasing the parameter $\tilde{\beta}$ in higher dimensions, but vice versa at low redshifts.
In Fig. \ref{Fig3}, we have plotted the evolution of the density abundance
$\Omega_{m}$, defined as
\begin{eqnarray}
\Omega_{m}&\equiv &\dfrac{2\rho_{m}}{n(n-1)H^{2}}=
\dfrac{2\tilde{\beta}(1+\tilde{\beta}-\Omega_{\Lambda
,0})(1+z)^n}{\left[ \sqrt{X(z)}-1\right]}. \label{OMmatter2}
\end{eqnarray}
As we can see from Fig. \ref{Fig3}, the matter density abundance with
different dimensions has the same behavior, i.e., all graphs are
reduced by decreasing $z$. In addition for higher dimensions ($n$
parameter), the density abundance drops faster.
Also we can see, the matter density abundance increases with
increasing the parameter $\tilde{\beta}$ in higher dimensions. \\
In a similar way, the evolution of the density abundance
$\Omega_{\Lambda}$ is given by
\begin{eqnarray}
\Omega_{\Lambda}&\equiv &\dfrac{2\Lambda}{n(n-1)H^{2}}=
\dfrac{2\tilde{\beta}\Omega_{\Lambda,0}}{\left[
\sqrt{X(z)}-1\right]}. \label{OMLambda2}
\end{eqnarray}

\begin{figure}[t]
\epsfxsize=7.5cm \centerline{\epsffile{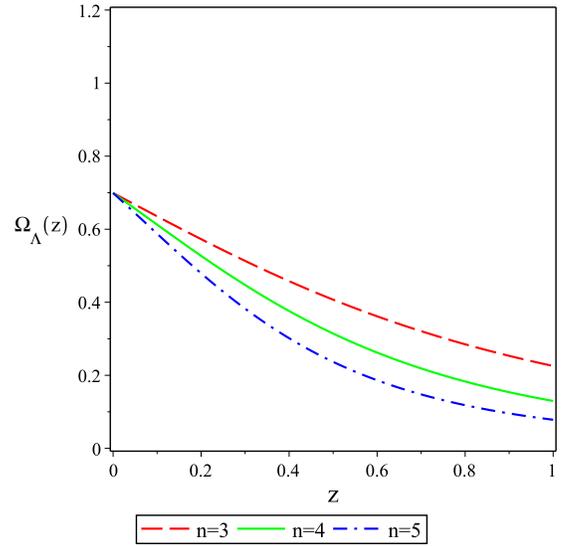}} \caption{The evolution of the DE density abundance as a
function of redshift z for different values of $n$, where we have taken $\tilde{\beta}=10^{-16}$.} \label{Fig4}
\end{figure}
\begin{figure}[t]
\epsfxsize=7.5cm \centerline{\epsffile{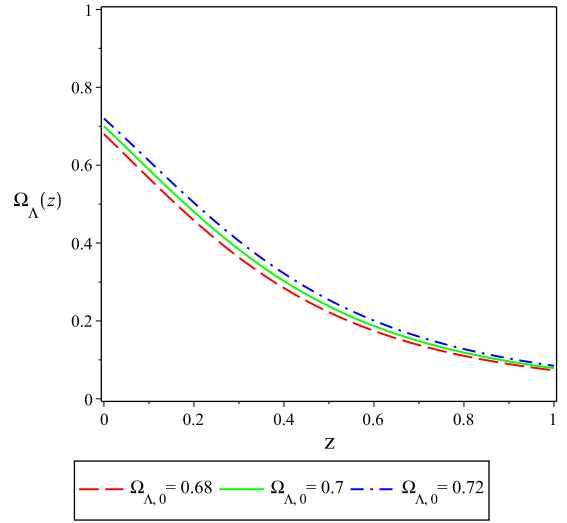}} \caption{The evolution of the DE density abundance as a
function of redshift z for different values of $\Omega_{\Lambda ,0}$, where we have taken $n=5,\tilde{\beta}=10^{-16}$.} \label{Fig5}
\end{figure}
\begin{figure}[t]
\epsfxsize=7.5cm \centerline{\epsffile{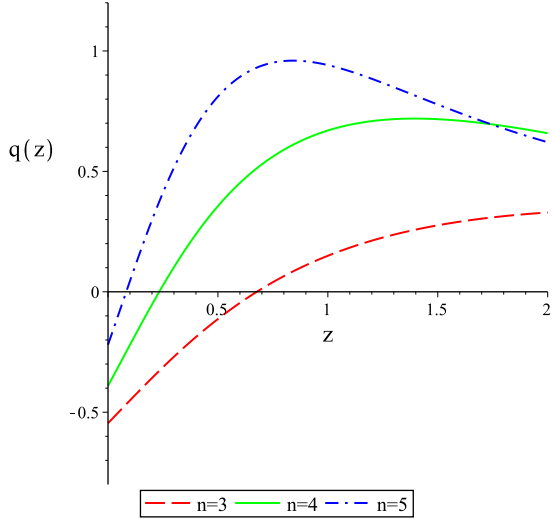}} \caption{The
behavior of the deceleration parameter $q(z)$ as a function of
redshift for different $n$. Here we have taken $\tilde{\beta}=10^{-16}$.} \label{Fig6}
\end{figure}

In Figs. \ref{Fig4}, and \ref{Fig5} we plot the evolution of the DE density abundance in
various dimensions and for different values of
$\Omega_{\Lambda,0}$ parameter. It is seen that the DE density
abundance $\Omega_{\Lambda}$ increases by decreasing $z$.
From Fig. \ref{Fig4}, we see that for a fixed value of redshift parameter $z$,
the value of the density abundance decreases with increasing the spacetime dimensions.\\
The deceleration parameter in terms of the redshift can be written
as
\begin{eqnarray}
&&q=-1-\dfrac{\dot{H}}{H^2}=-1+\dfrac{(1+z)}{H(z)}\dfrac{dH(z)}{dz} \nonumber\\
&&=-1+\dfrac{n \ \tilde{\beta}}{\sqrt{X(z)}}\dfrac{(1+z)^{n}(1+\tilde{\beta}-\Omega_{\Lambda
,0})}{\left[ \sqrt{X(z)}-1\right]}.
\label{deceleration}
\end{eqnarray}
We have plotted the behavior of the deceleration parameter $q(z)$
for different dimensions in Fig. \ref{Fig6}. We observe that for lower
dimensions the universe experiences a transition from a
decelerating phase ($q>0$) to an accelerating phase ($q<0$), at
redshift around $z=z_{tr}$. It is seen that $z_{tr}$ depends on
the spacetime dimension and decreases with increasing $n$.

\begin{figure}[t]
\epsfxsize=7.5cm \centerline{\epsffile{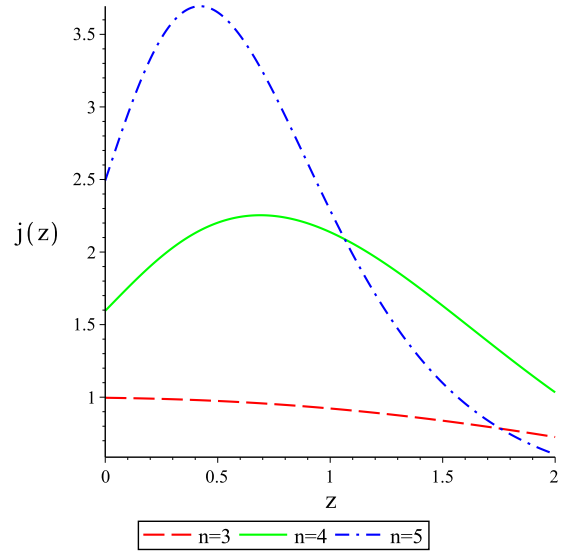}} \caption{The
evolution of jerk parameter with respect to redshift for different
values of $n$ parameter, Here we have taken $\tilde{\beta}=10^{-16}$.} \label{Fig7}
\end{figure}

Another quantity which is helpful in understanding the phase
transitions of the universe is called the \textit{jerk parameter}.
This is a dimensionless quantity obtained by taking the third
derivative of the scale factor with respect to the cosmic time,
provides a comparison between different DE models and the
$\Lambda$CDM $(j=1)$ model. The jerk parameter is defined as
\begin{equation}
j=\dfrac{1}{aH^{3}}\dfrac{d^{3}a}{dt^{3}}=q(2q+1)+(1+z)\dfrac{dq}{dz}.
\label{jerk}
\end{equation}
For the $\Lambda$CDM model, the value of $j$ is always unity. A
non-$\Lambda$CDM model occurs if there is any deviation from the
value of $j=1$. From Fig. \ref{Fig7} we observe that in the
context of GB cosmology, the jerk parameter is larger than
$\Lambda$CDM in higher dimensions.

%%%%%%%%%%%%%%%%%%%%%%%%%%%%%%%%%%%%%%%%%%%%%%%%%%%%%%%%%%%%%%%%%%%%%%%%%%%%%%%%%%%%%%%%%%%%%%%%%%%%%%%%%%%%%%%%%%%%%%%%%%%%
\section{Growth Of Perturbations in GB Cosmology\label{PGC}}
We consider a universe filled with pressureless matter,
$(p_{m}=0)$. In this case Eq. (\ref{continuty}) reads
\begin{equation}
\dot{\rho}_{m}+nH\rho_{m}=0,
 \label{continuty1}
\end{equation}
which has a solution of the form $\rho_{m}= \rho_{m,0}a^{-n}$,
where $\rho_{m,0}$ is the energy density at the present time. In
order to study the growth of perturbations, we consider a
spherically symmetric perturbed cloud of radius $a_{p}$, and with
a homogeneous energy density $\rho_{m}^{c}$. The SC model
describes a spherical region with a top-hat profile and uniform
density so that at any time $t$, we can write
$\rho_{m}^{c}(t)=\rho_{m}(t)+\delta\rho_{m}$ \cite{Ziaie2}. If
$\delta\rho_{m}>0$ this spherical region will eventually collapse
under its own gravitational force and if $\delta\rho_{m}<0$ it
will expand faster than the average Hubble expansion rate, thus
generating a void. In other words, $\delta\rho_{m}$ is positive in
overdense region and it is negative in underdense regions. In
fact, when the universe is in the matter dominated era, denser
regions expand slower than the entire universe. Therefore if their
density is enough large, they eventually collapse and create
gravitational constraints systems like clusters \cite{Ryden}.
Similar to Eq. (\ref{continuty1}), the conservation equation for
spherical perturbed region can be written as
\begin{equation}
\dot{\rho}_{m}^{c}+nh\rho_{m}^{c}=0, \label{continuty2}
\end{equation}
where $h=\dot{a}_{p}/a_{p}$ is the local expansion rate of the
spherical perturbed region of radius $a_p$ (subscript $p$ refers
to the perturbed). In order to study the evolution of
perturbations, we define a useful and dimensionless quantity
called density contrast as \cite{Ryden}
\begin{equation}
\delta_{m}=\dfrac{\rho_{m}^{c}}{\rho_{m}}-1=\dfrac{\delta\rho_{m}}{\rho_{m}},
\label{delta}
\end{equation}
where $\rho_{m}^{c}$ is the energy density of spherical perturbed
cloud and $\rho_{m}$ is the background density. Taking the
derivative of Eq.(\ref{delta}) with respect to the cosmic time and
using Eq.(\ref{continuty1}) and Eq.(\ref{continuty2}), we obtain
\begin{eqnarray}
&&\dot{\delta}_{m}=n(1+\delta_{m})(H-h),\label{deltadot} \\
&&\ddot{\delta}_{m}=n(\dot{H}-\dot{h})(1+\delta_{m})+
\frac{\dot{\delta}_{m}^{2}}{1+\delta_{m}}, \label{deltadubbledot}
\end{eqnarray}
where the dot denotes the derivative with respect to time.
Combining Eqs. (\ref{Fried1}), (\ref{Hz}), and
expanding $X(z)$ (only to the linear term of $\tilde{\beta}$),
 we arrive at
\begin{eqnarray}
\dfrac{\ddot{a}}{a}&=&\dfrac{(2-n)}{n(n-1)}\rho_{m}+\dfrac{4 \tilde{\alpha}}{n^{2}(n-1)}\rho_{m}^{2}+\dfrac{4\tilde{\alpha}(n-2)}{n^{2}(n-1)^{2}}\rho_{m}\Lambda\nonumber\\
&&+\dfrac{2}{n(n-1)}\Lambda -\dfrac{4 \tilde{\alpha}}{n^2(n-1)^{2}} \Lambda^{2}.
\label{adubbledot}
\end{eqnarray}
According to SC model, a homogeneous sphere of uniform density
with radius $a_{p}$ can itself be modeled using the same equations
that govern the evolution of the universe, with scale factor $a$
\cite{Peebles}. Therefore, we can write for the spherical
perturbed cloud  with radius $a_{p}$, an equation similar to
Eq.(\ref{adubbledot}), namely
\begin{eqnarray}
\dfrac{\ddot{a}_{p}}{a_{p}}&=&\dfrac{(2-n)}{n(n-1)}\rho_{m}^{c}+\dfrac{4 \tilde{\alpha}}{n^{2}(n-1)}(\rho_{m}^{c})^{2}+\dfrac{4\tilde{\alpha}(n-2)}{n^{2}(n-1)^{2}}\rho_{m}^{c}\Lambda\nonumber\\
&&+\dfrac{2}{n(n-1)}\Lambda -\dfrac{4
\tilde{\alpha}}{n^2(n-1)^{2}} \Lambda^{2}.
\label{apduubeldot-delta}
\end{eqnarray}
{In general, one may expect $\tilde{\alpha}$ differ inside and
outside of the spherical region. However, for simplicity here we
propose they are similar, namely
$\tilde{\alpha}^{c}=\tilde{\alpha}$. This can be easily understood
if we assume the role of the GB term at the early stages of the
universe is very weak, thus we can expect the same values for
$\tilde{\alpha}$ inside and outside of the spherical region. }

Combining Eqs. (\ref{delta}), (\ref{adubbledot}) and
(\ref{apduubeldot-delta}), yields
\begin{eqnarray}
\dot{H}-\dot{h}&=&\dfrac{(2-n)}{n(n-1)}\rho_{m} \delta_{m}-\dfrac{8 \tilde{\alpha}}{n^{2}(n-1)}\rho_{m}^{2}\delta_{m}\nonumber\\
&&-\dfrac{4\tilde{\alpha}(n-2)}{n^{2}(n-1)^{2}}\rho_{m}\Lambda\delta_{m}-H^{2}+h^{2},
\label{Hdot-hdot}
\end{eqnarray}
where we have expanded the second term and only kept the linear
term of $\delta_m$. This is due to the fact that we work in the
linear regime where $\delta_m<1$.

Substituting Eq. (\ref{Hdot-hdot}) into Eq. (\ref{deltadubbledot})
and using Eq. (\ref{deltadot}), we can find the second order
differential equation for the density contrast $\delta_m$ in the linear
regime as
\begin{eqnarray}
&&\ddot{\delta}_{m}+2H\dot{\delta}_{m}-\dfrac{(n-2)}{(n-1)}\rho_{m}\delta_{m}\nonumber\\
&&+\dfrac{8
\tilde{\alpha}}{n(n-1)}\rho_{m}^{2}\delta_{m}+\dfrac{4\tilde{\alpha}(n-2)}{n(n-1)^{2}}\rho_{m}\Lambda\delta_{m}=0.
\label{deltaduubeldot2}
\end{eqnarray}
In order to study the evolution of the density contrast $\delta_m$
in terms of the redshift parameter, $1+z=1/a$, we first replace
the time derivatives with the derivatives with respect to the
scale factor $a$. It is a matter of calculations to show that
\begin{equation}
\dot{\delta}_{m}=\delta_{m}^{\prime}aH, \quad \ddot{\delta}_{m}
=\delta_{m}^{\prime\prime}a^{2}H^{2}+a\left(H^{2}+\dot{H}\right)\delta^{\prime}_{m},
\label{prim}
\end{equation}
where the prime stands for the derivative respect to the scale
factor $a$. Using Eqs.(\ref{Hz}) and (\ref{adubbledot}), we get
\begin{eqnarray}
\dot{H}&=&-\dfrac{1}{n-1}\rho_{m}+\dfrac{4\tilde{\alpha}\Lambda}{n^{2}(n-1)^{2}}\nonumber\\
&&+\dfrac{4\tilde{\alpha}}{n(n-1)^{2}}\rho_{m}^{2}+\dfrac{4\tilde{\alpha}}{n^{2}(n-1)}\Lambda \rho_{m}.
\label{Hdot}
\end{eqnarray}
Therefore Eq. (\ref{deltaduubeldot2}), after using Eqs.
(\ref{prim}) and (\ref{Hdot}), can be written as
\begin{eqnarray}\label{deltafora}
&&\delta^{\prime\prime}_{m}+\dfrac{3}{2a}\delta^{\prime}_{m}-\dfrac{(n^{2}-4\tilde{\alpha}\Lambda)}{n^{2}(n-1)}\dfrac{\rho_{m}}{aH^{2}}\delta^{\prime}_{m}+\dfrac{4\tilde{\alpha}\Lambda}{n^{2}(n-1)^{2}}\dfrac{1}{aH^{2}}\delta^{\prime}_{m}\nonumber\\
&&+\dfrac{4\tilde{\alpha}}{n(n-1)^{2}}\dfrac{\rho_{m}^{2}}{aH^{2}}\delta^{\prime}_{m}-\dfrac{(n^{2}-n-4\tilde{\alpha}\Lambda)(n-2)}{n(n-1)^{2}}\dfrac{\rho_{m}}{a^{2}H^{2}}\delta_{m}\nonumber\\
&&+\dfrac{8\tilde{\alpha}}{n(n-1)}\dfrac{\rho_{m}^{2}}{a^{2}H^{2}}\delta_{m}=0.
\end{eqnarray}
Since we are working in the linear regime, we neglect
$O(\delta_{m}^2)$ and $ O({\delta^{\prime}_m}^2)$. Combining
Eqs.(\ref{Hz}) and (\ref{deltafora}), we arrive at
\begin{eqnarray}
&&\delta^{\prime\prime}_{m}+\dfrac{3}{a}\delta^{\prime}_{m}-\dfrac{1}{a\ \Gamma}\Bigg{\{} \dfrac{(n^{2}-4\tilde{\alpha}\Lambda)\rho_{m}}{{2n}}-\dfrac{2\tilde{\alpha}\Lambda}{n(n-1)}\nonumber\\
&&-\dfrac{2\tilde{\alpha}\rho_{m}^{2}}{(n-1)}\Bigg{\}}\delta^{\prime}_{m}-\dfrac{1}{a^{2}\ \Gamma}\Bigg{\{} \dfrac{(n^{2}-n-4\tilde{\alpha}\Lambda)(n-2)\rho_{m}}{{2(n-1)}}\nonumber\\
&&-(4\tilde{\alpha}\rho_{m}^{2})\Bigg{\}}\delta_{m}=0.
\nonumber\\
\label{betadelta}
\end{eqnarray}
where
\begin{equation}
\Gamma=(\rho_{m}+\Lambda) \left[ 1-\dfrac{2\tilde{\alpha}}{n(n-1)}(\rho_{m}+\Lambda) \right] .
\label{Gamma}
\end{equation}
It should be noted that in the limiting case where $n =3$ and
$\Lambda=0$, Eq. (\ref{betadelta}) reduces to
\begin{eqnarray}\label{deltagr}
&&\delta^{\prime\prime}_{m}+\dfrac{3}{2a}\delta^{\prime}_{m}-\frac{3\delta_{m}}{2a^{2}}=0,
\end{eqnarray}
which is the result obtained in standard cosmology \cite{Abramo}.
In other words, in the absence of the GB term, the perturbed
equation for the density contrast, $\delta_m$, in the linear
regime, coincides with the corresponding equation in standard
cosmology.

\begin{figure}[t]
\epsfxsize=7.5cm \centerline{\epsffile{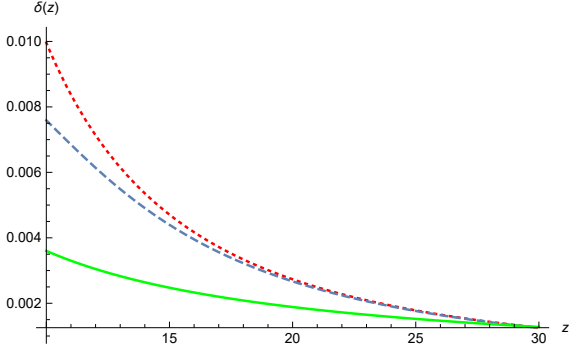}}
\caption{The 
evolution of the matter density contrast for different values of
$(\tilde{\beta}, n)$, where dotted-line for $(10^{-5}, 4)$, 
dashed-line for $(0, 4)$ and solid-line for $(0,3 (\Lambda CDM))$. 
We have chosen $\delta_{m}(z_{i})=0.0001$ , $z_{i}
= 400$.} \label{Fig8}
\end{figure}

Since Eq.(\ref{betadelta}) has no analytic solution, we
numerically plot, in Fig. \ref{Fig8}, the matter density contrast for
different values of $\tilde{\beta}$ parameter in various dimensions. We
observe that in the framework of GB gravity in five dimensions,
the density contrast of matter starts growing from its initial
value and, as the universe expands, the matter density contrast
grows up faster and deviates from the standard model. Indeed, the
growth of matter perturbations in five spacetime dimensions is
faster comparing to the standard four dimensional cosmology. This
means that in a universe with extra dimensions, the structures
forms sooner. Besides, with increasing $\alpha$ ($\tilde{\beta}$)
parameter, the matter perturbations grows faster. This is an
expected result, since the higher order GB correction terms
increase the strength of the gravity and thus support the growth
of perturbations

\begin{figure}[t]
\epsfxsize=7.5cm \centerline{\epsffile{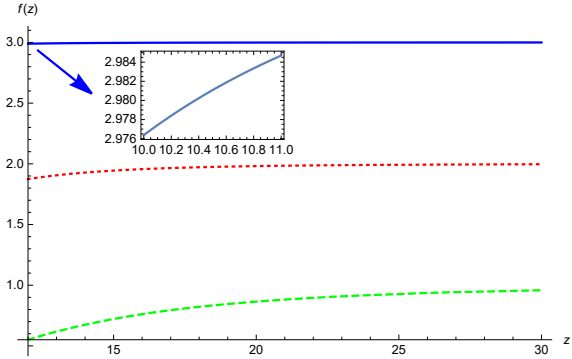}} \caption{The
evolution of the growth function with $\tilde{\beta}=10^{-6}$ and for different values of $n$
parameter, where dashed-line for $n=3$, dotted-line for $n=4$
and solid-line for $n=5$.}
\label{fig9}
\end{figure}

We can investigate the growth rate of matter perturbations which
is given by the growth function as \cite{Peebles}
\begin{equation}\label{ growth rate}
f(a)=\dfrac{dlnD}{dlna},\      \     \     \    \
D(a)=\dfrac{\delta_{m}(a)}{\delta_{m}(a=1)}. 
\end{equation}
Let us note that in the absence of the GB term ($\tilde{\alpha}=
0$), the growth function is a constant of unity. In Fig. \ref{fig9} we
have plotted the growth function in terms of the redshift
parameter. We observe that in the framework of GB gravity, the
growth function in higher dimensions grows faster than
four-dimensional GB and the growth function increases with increasing
$n$.

\begin{figure}[t]
\epsfxsize=6.8cm \centerline{\epsffile{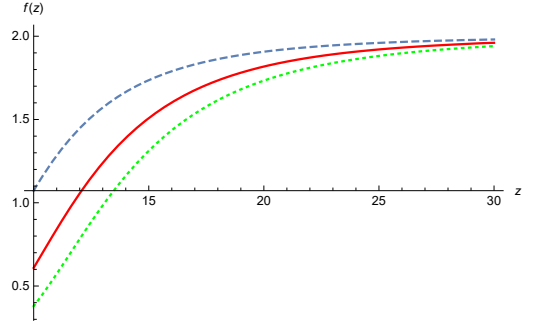}} \caption{The
evolution of the growth function for different values of $\tilde{\beta}$
parameter in 5D, where dotted-line for $\tilde{\beta}=10^{-5}$, solid-line for $\tilde{\beta}=2\times 10^{-5}$ and
dashed-line for $\tilde{\beta}=3\times 10^{-5}$. } \label{fig10}
\end{figure}
From Fig. \ref{fig10}, we can see that the growth function
increases with increasing $\tilde{\beta}$ parameter in higher-dimensional GB cosmology.\\

 %%%%%%%%%%%%%%%%%%%%%%%%%%%%%%%%%%
\section{Halo abundance in GB cosmology}\label{MN}
This section is devoted to study the distribution of the number
density of collapsed objects of a given mass range in the
framework of GB cosmology. The collapsed objects which, in
essence, are the main source of large-scale structure formation of
the universe are called the DM halos. Besides, the baryonic matter
due to the gravitational attraction follows the DM distribution.
In this way, tracing the distribution of DM haloes becomes
possible by looking at the distribution of galaxy clusters. In
order to investigate the number density of distribution of the
collapsed objects or the galaxy clusters along the redshift,  a
semi-analytic approach known as the Press-Schechter formalism is
commonly employed \cite{William}. Usually, the matter density
field in the mathematical formulations of the halo mass function
should be enjoyed the Gaussian distribution.

The comoving number density of the gravitationally collapsed
objects (equivalent to galaxy clusters) at a certain redshift $z$,
with mass from $M$ to $M + dM$, is given by the following
analytical formula \cite{Liberato}
\begin{equation}
\dfrac{dn(M,z)}{dM}=-\dfrac{\rho_{m,0}}{M}\dfrac{d\ln
\sigma(M,z)}{dM} f(\sigma(M,z)),
 \label{massfunction}
\end{equation}
where $\rho_{m,0}$, $\sigma(M, z)$ and $f(\sigma)$, respectively,
denote the present matter mean density of the universe, the rms of
density fluctuation in a sphere of radius $r$ surrounding a mass
$M$, and the mathematical mass function proposed by Press and
Schechter \cite{William}, as follow
\begin{equation}
f_{PS}(\sigma)=\sqrt{\dfrac{2}{\pi}} \dfrac{\delta_{c}(z)}{\sigma(M,z)}\exp\left[ -\dfrac{\delta_{c}^{2}(z)}{2\sigma^{2}(M,z)}\right].
\label{fsigma0}
\end{equation}
Subscript ''PS'', refers to Press and Schechter. Note that
$\delta_{c}(z)$ in the mass function above is the critical density
contrast above which structures collapse. By serving the linerised
growth factor $D(z) = \delta_{m}(z)/\delta_{m}(z=0)$, as well as
the rms of density fluctuation at a fixed length $r_{8} =
8h^{-1}Mpc$, one can express $\sigma(M, z)$ as
\begin{equation}
\sigma(z,M)= \sigma(0,M_{8})\left(\dfrac{M}{M_{8}}
\right)^{-\gamma/3} D(z),
 \label{sigmazm}
\end{equation}
where the index $\gamma$ reads as \cite{Mukherjee,Mukherjee2}
\begin{equation}
 \gamma=(0.3\Omega_{m,0}h+0.2)\left[ 2.92+\dfrac{1}{3} \log\left(  \dfrac{M}{M_{8}}\right) \right],
 \label{gamma}
\end{equation}
and $M_{8}=6\times10^{14}\Omega_{M}^{(0)}h^{-1}M_{\bigodot}$ is
the mass inside a sphere of radius $r_{8}$ ($M_{\bigodot}$ is the
solar mass) \cite{Viana}.

\begin{figure}[t]
\epsfxsize=7.5cm \centerline{\epsffile{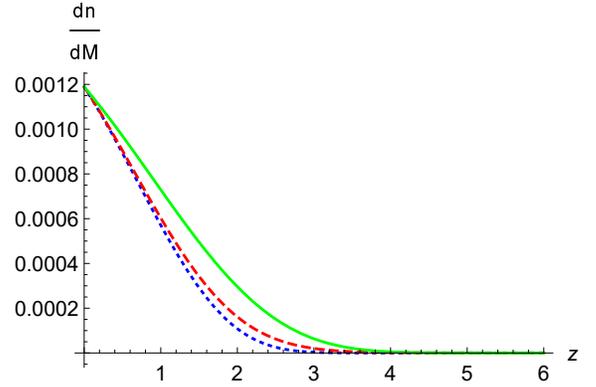}} \caption{The
evolution of mass function for objects with mass
$M=10^{13}(h^{-1}M_{\bigodot})$, and different values of $(\tilde{\beta},n)$, 
where dotted-line for $(10^{-7},4)$, dashed-line for $ (10^{-5},4)$
and solid-line for $(10^{-5},5)$.}
\label{fig11}
\end{figure}

In Fig. \ref{fig11}, we display the redshift evolution of mass function,
$dn/dM (1/Mpc^3)$, of objects with mass $10^{13}
h^{-1}M_{\bigodot}$ for different values of $\tilde{\beta}$ in various
dimensions. This figures explicitly shows that in each dimension,
as the parameter $\tilde{\beta}$ increases, the halo abundance grows
rapidly at lower redshifts. Besides, for a given GB parameter,the
mass function with increasing the spacetime dimension. In other
words, the halo abundance is formed later in lower dimensions.

To sum up, in this section we saw that the GB parameter as well as
the extra dimensions influence the evolution of the mass function
of the dark matter halos, therefore comparing to standard
$\Lambda$CDM, our model can predict different results for
clustering of galaxy clusters.\\

%%%%%%%%%%%%%%%%%%%%%%%%%%%%%%%%%%%%%%%%%%%%%%%%%%%
\section{Conclusions and discussions\label{CD}}
The growth of perturbations at the early stages of the universe
and the formation of galaxies and structures, due to the
gravitational collapse, is still an open question in modern
cosmology. It is instructive to explore how this phenomena occurs
in different gravity theories. In the present work, we have
explored the gravitational collapse of matter at the early
universe when the higher order corrections on the gravity side are
present in the action. We have investigated the evolution of the
matter perturbations in the context of GB gravity in a flat
universe filled with DM and DE  (cosmological constant) for
different values of the model parameters. We have employed the SC
formalism in order to examine the perturbations and worked in the
linear regime for the matter density contrasts $\delta_m$ as well
as the GB coupling ${\tilde{\beta}}=(n-2)(n-3)\alpha H_0^2$. We observe that the
density contrast has similar behavior for different values of
$\tilde{\beta}$ parameter; that is, it starts from its initial value and
then the growth of perturbations increases with increasing $\tilde{\beta}$
parameter, which reveals the influences of $\tilde{\beta}$ in GB
cosmology. Interestingly enough, we found that the growth of
perturbations increases with increasing $\alpha$. This is an
expected result, since the higher order GB correction terms
increase the strength of the gravity and thus support the growth
of perturbations. Besides, the matter density contrast $\delta_m$
in small redshifts grows faster in higher dimensions. Physically,
this means that the structures form faster in a universe with
extra dimensional spacetime. We have also studied the evolution of
the density parameters. We observed that the evolution of the
matter density abundance ($\Omega_{m}$) and DE density abundance
($\Omega_{\Lambda}$) for different values of $\tilde{\beta}$ decrease at
low redshifts. We found out that the density abundance of matter,
consist of baryonic and DM, drops slower for smaller values of
$\tilde{\beta}$ in higher dimensional GB cosmology. From the evolution of
the deceleration parameter, we see that the universe experiences a
phase transition from decelerated phase to an accelerated phase
around redshift $z_{tr}$. We saw that $z_{tr}$ is smaller in
higher dimensions, which means that our universe experiences this
phase transition later in higher dimensions. Also we observed that
in the framework of GB gravity, the growth function increases in
higher dimensions, and also increases with increasing $\tilde{\beta}$
parameter in higher-dimensional GB cosmology. Also we found that, the 
abundance of halo dark matter in higher dimensions start to grow in high reshifts,
so we can conclude that large scale structure(LSS) grows faster in this universe. 

%%%%%%%%%%%%%%%%%%%%%%%%%%%%%%%%%%%%%%%%%%%%%%%%%%%%%%%%%%%%%%%%%%%%%%%%%%%%%%%%%%%%%%%
\acknowledgments{We are grateful to M. Khodadi for useful 
discussion and valuable comments. }
%%%%%%%%%%%%%%%%%%%%%%%%%%%%%%%%%%%%%%%%%%%%%%%%%%%%%%%%%%%%%%%%%%%%%%%%%%%%%%%%%%%%%%%%%%%%%%%%%%%%%%%%%%%%%%%%%%%%%%%%%%%%%%%%%%%%%%%%%%%%%%%%%%%%%%%%%

\end{document}